\documentstyle[aps,twocolumn,prl]{revtex}
\begin{document}
\draft
\twocolumn[\hsize\textwidth\columnwidth\hsize\csname @twocolumnfalse\endcsname
\title{Decoherence and Quantum Fluctuations}
\author{P. Mohanty
        and R. A. Webb     \\
        Center for Superconductivity Research,\\
        Department of Physics, University of Maryland, College Park, 
	Maryland 20742}
\maketitle

\begin{abstract}

We show that the zero-point fluctuations of the intrinsic
electromagnetic environment limit the phase coherence 
time in all mesoscopic systems
at low temperatures. 
We derive this quantum noise limited 
dephasing time and its temperature dependence in the crossover
to the thermal regime. Our results agree well with most 
experiments in 1D systems.

\end{abstract}
]

\makeatletter
\global\@specialpagefalse
\let\@evenhead\@oddhead
\makeatother

\par

The importance of decoherence in the quantum to classical transition 
is well known\cite{zurek:physicstoday}.
There are many examples where a quantum system
undergoes environmentally induced decoherence 
such as the electron wavefunction
in a low dimensional solid, an atom trapped in a quantum optical system, and
an isolated quantum system coupled to a measuring apparatus.
It has also been suggested that the decoherence which must have occured
between the quantum and classical regimes during the inflationary
era of the Universe was driven by quantum
fluctuations\cite{zurek:physicstoday}.
Under certain conditions,
quantum zero-point fluctuations
of the electromagnetic environment can also cause decoherence, even at finite 
temperatures, and dominate over the thermal contributions\cite{Hu}.
These zero-point fluctuations which persist even
in vacuum are responsible for the
Lamb shift, the natural linewidths of energy levels, the Casimir force,
and the broadening of resonance lines in neutron scattering
in solids\cite{milonni}. In addition, it has recently 
been shown experimentally
that these zero-point fluctuations might play an important role in the
interference effects studied in mesoscopic systems\cite{us}.

\par

In quantum theory, the probabilistic amplitudes are added 
up using the linear superposition principle. The net 
probability contains along with the classical probability,
the interference or the off-diagonal
terms of the corresponding density matrix which are characteristics
of the quantum nature of the system\cite{feynman}. The natural
way to define decoherence quantitatively is to obtain a time scale
$\tau_\phi$ over which interference effects are suppressed, or the
relative phases in different Feynman
paths of the electron wavefunction are randomized. This is 
essentially the time over which an electron
maintains its phase memory. Normally, phase randomization of the electron 
wavefunction in a mesoscopic system occurs 
due to inelastic scattering effects
such as electron-electron, electron-phonon and 
magnetic impurity interactions\cite{altshuler,cs,sai,loss}, all of which
can be considered environmental for the electron under study.
We suggest, however, that zero-point fluctuations 
also contribute
to decoherence of the electron wavefunction 
resulting in physically observable
effects, such as the suppression of interference phenomena in mesoscopic
systems.

\par

In this paper we show how zero-point fluctuations can cause
dephasing in a  mesoscopic system. For  
a 1D quantum wire, we present two approaches for
obtaining the
zero-point limited dephasing time $\tau_{0}$ at low temperatures
which are in good agreement with the measured saturation
values of $\tau_\phi$ found in many experiments. 
We find that $\tau_{0}$ essentially depends on the
diffusion constant $D$ and the resistance per unit length of the
sample under study. We also derive the functional
form of   
$\tau_\phi(T)$ which is 
valid from the zero    
temperature quantum regime into the high temperature
regime, and accurately describes the observed behavior in the
quantum to classical crossover regime.
  
\par

An  example of the problem  we are
trying to understand is the temperature dependence of the
phase coherence time $\tau_\phi$ in a quasi-1D wire  
illustrated in Fig. 1. 
The gold wire has a resistance of 271 $\Omega$, it is 
207 $\mu$m long, 0.11 $\mu$m wide, 0.06 $\mu$m thick, and has
a classical diffusion constant D = 0.068 m$^2$/s. At the lowest
temperature displayed, the phase coherence length $L_\phi$
determined from standard weak localization measurements\cite{altshuler}
is 15.7 $\mu$m and the phase coherence time $\tau_\phi = L_\phi^2/D
= 3.6$ ns.
Contrary to every theoretical prediction,
$\tau_\phi$ is essentially temperature
independent at low temperatures, 
and this type of behavior is seen in every
experiment on 1D wires and 2D films, however the temperature at which
the saturation behavior starts varies from 20 mK to 10 K depending
on the system under investigation\cite{us,lin,pooke,hiramoto,bergmann}. It has been shown that this
saturation behavior is not due to magnetic impurity scattering, or
heating of the electrons due to noise or excess power dissipation from    
the external environment\cite{us}. The solid line drawn through the data was
experimentally determined to be

\begin{equation}
\tau_\phi = \tau_0 \tanh\big[ \alpha\pi^2\sqrt{{\hbar \over {\tau_0 k_B T}}}
\big],
\end{equation} 
\noindent
where $\tau_0$ is the low temperature saturation value of $\tau_\phi$.
Eq. 1 was shown to fit most 1D experiments on mesoscopic metal wires
with constant $\alpha$ only varying from 0.6 to 1.1 \cite{us}.

\par

The functional form of Eq. 1 suggests that the zero-point
fluctuations in the electromagnetic environment 
\begin{figure}
 \vbox to 5cm {\vss\hbox to 7cm
 {\hss\
   {\includegraphics{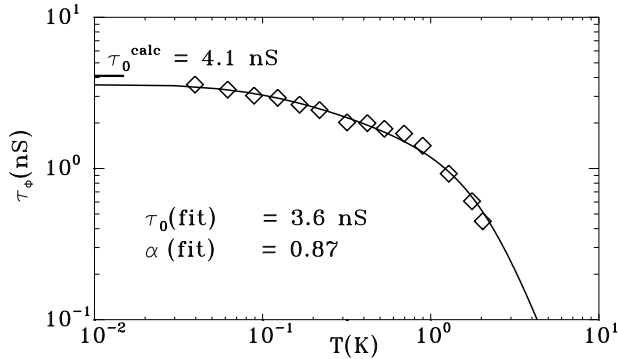}
   }
  \hss}
 }
\caption{Temperature dependence of $\tau_\phi$ for a quasi-1D gold wire.
The solid line is a fit to Eq. 1 with phonons.
Our predicted $\tau_0^{calc}$ from Eq. 6 is shown
on the plot.}
\end{figure}
\noindent
are responsible for the decoherence of $\tau_\phi$ found at
low temperatures in
many experiments in mesoscopics. The zero-point field fluctuations
can also be interpreted as a source field effect which 
means that one can equivalently study a quantum system
coupled to a fluctuating electromagnetic source field.
The generalized spectral density function which includes the
zero-point fluctuations of the electromagnetic
environment\cite{callenwelton,landau} is

\begin{equation}
S(\omega) = \hbar\omega({1\over 2}+{1\over{e^{\hbar\omega/{k_BT}}-1}}).
\end{equation}
\noindent
At low temperatures, the zero-point fluctuations 
dominate and $S(\omega)$ becomes      
temperature independent. The quantum system equilibrates
with a quantum statistical system 
described by the zero-point fluctuations.
At
high temperatures, $S(\omega) \rightarrow k_BT$, thus losing
the quantum signature $\hbar$ of the electromagnetic
fluctuations and $S(\omega)$ is dominated by 
the thermal fluctuations. The quantum system in this case
relaxes into equilibrium with a thermal bath at temperature $T$. 
In the intermediate regime,
the full form for $S(\omega)$ must be retained in order
to describe the transition between these two regimes.

\par

We make a brief remark about what happens at low 
temperatures. When the system or a particle is coupled
to an environment which possesses 
many degrees of freedom, even for weak couplings,
the Hamiltonian for the entire system cannot be exactly diagonalized
in the eigenbasis of the particle. The lack
of a diagonal basis is the reason that 
the system can only evolve to mixed
states even at zero temperature. 
One may construct, or in principle it is 
possible to realize in a physical system, an orthogonal
basis for the full Hamiltonian in which
the system may continue to stay in a
well defined stationary state. The choice of
an initial state is also as important as the interaction.
Usually an eigenstate of the full Hamiltonian is
not easily prepared in realistic systems. 
Hence, the possibility of a transient behavior
of the system as it
comes into contact with a temperature bath cannot
be ruled out.

\par

To quantitatively estimate dephasing 
it suffices to evaluate the phase change of the electron
wavefunction in a fluctuating
electromagnetic field. It is well known that electron-electron 
interaction\cite{altshuler,cs} at low temperatures 
may be studied by equivalently
calculating the fluctuations in the electromagnetic field 
due to the other electrons which
serve as the environment for the electron under study. The time
dependence of this electromagnetic field causes decoherence.
This
is also equivalent to studying the changes induced in the 
states of the environment, 
as first pointed out by Chakravarty and Schmid\cite{cs}.
The equivalence, which is merely an artifact of the 
fluctuation-dissipation theorem, was very nicely elucidated
later by Stern, Aharonov and Imry\cite{sai}. 

\par

We express the acquired phase $\phi$ in the electron wavefunction
as the time integral
of the interaction potential
of the type
$V(x(t),t)={\dot {\bf x}}{\cdot}{\bf A}$ over the period
of
interaction. The quantity of interest is the average
of the phase factor $\langle e^{i\phi}\rangle$.
Here, $\langle \rangle$ implies averaging
due to disorder and field fluctuations which include
both quantum and thermal parts. 
$\langle e^{i\phi}\rangle$ can be approximated
as $e^{-{1\over 2}\langle\phi^2\rangle}$, because the mean value 
$\langle \exp(\alpha_i x_i)\rangle$
is given by $\exp({1\over 2} \alpha_i \alpha_j\langle x_ix_j\rangle)$
where $\alpha_i$ are constants and the $x_i$ are fluctuating quantities
subject to a Gaussian distribution\cite{landau}.
The suppression of the intensity 
of interference by dephasing
is of the form $e^{-\langle\phi^2\rangle/2}$ 
which, for $\langle\phi^2\rangle \sim t$, \cite{cs,sai}
becomes $e^{-t/\tau_\phi}$.
Generally $\langle\phi^2\rangle$ is not a linear function of time.
One has to estimate the evolution of the off-diagonal terms of
the reduced density matrix for the particle. The characteristic
time over which $\langle\phi^2\rangle$ becomes of the order of unity can be 
defined as the corresponding dephasing time.
We evaluate 
$\langle\phi^2\rangle$ 
by calculating the two time
integral of the function
${\dot {\bf x}_\alpha}{\dot {\bf x}_\beta}\langle {A}_\alpha{A}_\beta\rangle$
\cite{altshuler,cs,sai}.  
Using the fluctuation-dissipation theorem in a gauge where
the scalar potential is chosen to be zero one obtains the
two-point correlation function for the vector potential
$\langle A_{\alpha}(x(t),t)A_{\beta}(x(t^\prime),t^\prime)\rangle$ 
in the corresponding Fourier space $({\bf k},\omega)$ 
\cite{sai,landau}

\begin{equation}
\langle A_\alpha A_\beta\rangle_{k,\omega} = {\coth(\hbar\omega/2k_B T) \over \sigma\omega}
[{k_\alpha k_\beta \over k^2}]
\end{equation}
\noindent
where $\sigma$ is the conductance. This is valid for conducting systems at
all temperatures where the fluctuations are strongly suppressed by
the skin effect\cite{altshuler}. 
Following an approach similar to Chakravarty and Schmid\cite{cs},
the dephasing time $\tau_\phi$ can be expressed using Eq. 3 
by summing over all allowed frequencies and wave vectors.

\begin{eqnarray}
{1\over \tau_\phi}={e^2 \over \sigma\hbar^2}\int du\int {d{\bf k} \over (2\pi)^2}
\int {d\omega \over {2\pi}} 
\hbar\omega \coth({\hbar\omega \over {2k_BT}}) \nonumber \\
 \times k^{-2} \exp(-Dk^2|u|-i\omega u).
\end{eqnarray}
\noindent
In this expression, $1/\sigma$ for a quasi-1D wire is the resistance 
of a phase coherent length, $R_\phi= RL_\phi/L$.  
In a quasi-1D system, for any two interfering paths the wave vector
${\bf k}$ must be a two dimensional vector. 

\par

Previous attempts to incorporate zero-point fluctuations
only considered the motion of impurity ions\cite{kumar}.
Our approach differs from all others in that to calculate the low
temperature saturation value of phase coherence time $\tau_0$
we consider only the zero-point fluctuations of the intrinsic
electromagnetic field . 
At low temperatures or high frequencies, $\omega >> k_BT/\hbar$,
the contribution to
dephasing is dominated by the quantum part of the field fluctuations
which diverges linearly for large $\omega$. As in previous 
theories\cite{altshuler,cs,sai}, one must introduce a phenomenological
upper cutoff to obtain any
non-divergent fluctuation dependent physical quantity. In the case
of electrons in a quasi-1D disordered system, the fluctuations
must be able to ``couple to the particle displacement''  
\cite{loss,landau,zurekunruh} in order 
to contribute effectively to decoherence. The maximum frequency
of the fluctuations that one should retain is no longer $k_BT/\hbar$,
but becomes the average classical energy
of the particle. For a ballistic sample this cutoff would then be the
Fermi energy, $E_F$, but for a diffusive sample it becomes $m^\ast v_D^2/2$,
where $m^\ast$ is the effective mass,
$v_D$ is the drift velocity of the particle 
given by $v_Fl_e/L$. $l_e$ is the elastic 
mean free path, and for a phase coherent volume, $L$ is essentially
the phase coherent length, $L_\phi$. Such a choice of the upper
cutoff is also natural in other models of decoherence where
the inverse of the collision timescale  is typically taken
as the cutoff\cite{zurekunruh}. 
Thus frequencies much higher than $k_BT$ contribute to dephasing.
In the problem at hand, $|\omega|$ in Eq. 4 is also bounded
at the bottom by $hD/L_\phi^2$. The electron
traverses
an average distance of $L_\phi$ before losing phase
coherence.  Therefore the maximum wavelength 
which can couple to the particle  is $k^{-1} \sim L_\phi$.
Wavelengths longer than $L_\phi$ do not contribute 
to dephasing. The lower cutoff
in $|\omega|$ is given by $Dk^2 \simeq D/L_\phi^2 = 1/\tau_\phi$. 

\par

Extending the $\omega$ integral over the range 
$h/\tau_\phi \le |\hbar\omega| \le m^\ast v_D^2/2$, and 
taking the low temperature limit of the spectral density
function $S(\omega)\rightarrow \hbar\omega/2$, the
zero-point energy for the mode $\omega$, one obtains from Eq. 4 

\begin{equation}
{1\over \tau_0} = {e^2 R_\phi \over 2\pi\hbar^2}
\int_{hD/L_\phi^2}^{m^\ast v_D^2/2} d(\hbar\omega).
\end{equation} 
\noindent
Except in very diffusive systems, the lower cutoff is 
small compared to the upper cutoff.  
Using
$L_\phi(T\rightarrow 0) = \sqrt{D\tau_0}$, 
the 1D zero-point phase coherence time $\tau_0$ is given by

\begin{equation} 
{1\over \tau_0} = \big( {{e^2 d^2Rm^\ast D^{3/2} } \over
			 {4\pi\hbar^2 L}}\big)^{2}.
\end{equation}
\noindent
We have used $R/L=R_\phi/L_\phi$ and $D=v_F l_e/d$, where $D$
is the classical diffusion constant in $d$ dimensions.
Using the sample parameters given earlier, for the 1D wire
displayed in Fig. 1, we find $\tau_0 \simeq 4.1$ nS, which
is in good agreement with the measured saturation value of
$3.6$ nS.
As shown in Ref. \cite{us}, this expression also accurately predicts
the value of the low temperature saturation phase coherence time $\tau_0$
found in most published experiments on diffusive 1D wires
including semiconductors. For a high mobility 2DEG wire
with system size $L \simeq L_\phi$, from Eq. 4 
one easily obtains an expression for the zero-point dephasing
time $1/\tau_0 = (m^\ast D/4h)(\Delta/\hbar)$ by using the
Einstein relation $\sigma = e^2 N(0)D$, where  $\Delta = 2/N(0)A$.
$N(0)$ is the density of states at $E_F$, and $A$ is the
phase coherent area of the sample. 

\par

In the high temperature regime, Eq. 2 reduces to $S(\omega)= k_BT$
\cite{altshuler,cs,sai,loss}.
The upper cutoff for $\omega$ can then be taken as $k_BT$
since the thermal fluctuations dominate with a thermal correlation
time $\hbar/k_BT < \tau_0$. 
The crossover between the two regimes is very important
to understand. Because of the finite frequency cutoffs in Eq. 5, 
in a phenomenological model one can approximate the integrand in Eq. 4 
by the
spectral density function for a single average energy
mode, $\hbar\langle\omega\rangle \equiv \langle E\rangle$.
The expression for the temperature dependence of
$\tau_\phi$ with a zero-point saturation time $\tau_0$ 
becomes

\begin{equation}
1/\tau_\phi = (1/\tau_0) \coth(\langle E\rangle/2k_BT).
\end{equation}
\noindent
The above equation can be 
obtained from the rigorous solution of a Pauli Master equation
for the decohering two level system coupled to a thermal
bath\cite{agarwal} for a single mode $\langle E\rangle/\hbar$. 
We interpret  $\langle E\rangle$ to be the Thouless energy for
a particle that diffuses over a volume defined by 
two diffusion length scales, the
zero-point diffusion length 
$L_0 = \sqrt{D\tau_0}$, and 
the thermal diffusion length $L_T = \sqrt{\hbar D/k_BT}$.
The corresponding
Thouless energy is then given by $\langle E\rangle=\pi hD/L_0 L_T$.     
Eq. 4 can then be written as 
$\tau_\phi = \tau_0 \tanh [\alpha\pi^2\sqrt{\hbar \over \tau_0k_BT}]$,
where $\alpha$ is a constant of order unity which takes into account that
our approximation for $\langle E\rangle$ is only good to
first order. This is the same as Eq. 1 which was experimentally
deduced\cite{us} from studies of the dephasing time
in 1D wires. Interestingly, the temperature dependence
of $\tau_\phi$ given by Eq. 1 at high temperatures 
reduces to that given by electron-electron interactions
with large energy transfers\cite{altshuler}.
As shown by the solid line in Fig. 1, Eq. 1 correctly
describes the temperature dependence of the dephasing time
for this sample as well as most 1D mesoscopic wires published
to date once the phonon contribution to dephasing 
at high temperature is included by $1/\tau_\phi \rightarrow
1/\tau_\phi + 1/\tau_{ep}$, where $\tau_{ep}$ is the scattering   
time due to phonons. For 2D metal films, it has been shown
recently\cite{us} that Eq. 1 also correctly describes the
temperature dependence of $\tau_\phi$ with the constant $\alpha$         
reduced by a factor of $\pi$. For 1D semiconductor wires with
$L_T \gg w$, the width of the sample, Eq. 1 becomes $\tau_\phi
= \tau_0\tanh[(\alpha\hbar/k_BTw)\sqrt{D/\tau_0}]$, with
$\tau_0$ given by Eq. 6, and quantitatively describes the
complete behavior observed in many published experiments.
Note that at high temperatures for this case, $\tau_\phi \propto
1/k_BT$ and not to $(1/k_BT)^{1/2}$ as is the case for most
metal wires.

\par

It is interesting to note Eq. 6 for the intrinsic dephasing
time for mesoscopic wires $\tau_0$ could have also been
derived from a combination of the Aharonov-Bohm effect
and the fluctuation-dissipation theorem. One very general way
to write the fluctuation-dissipation theorem\cite{landau}
is to relate the mean square fluctuations, $\langle F^2\rangle$, of any
generalized force of a system in thermodynamic equilibrium
with a parameter, $R(\omega)$ which characterizes an irreversible
process,                      

\begin{equation}
\langle F^2\rangle={2\over \pi}\int R(\omega)S(\omega) d\omega,
\end{equation}
\noindent
where $S(\omega)$ is given by Eq. 2. 
Voltage across a 
resistive element is an example of a generalized force 
and Eq. 8 reduces to the standard Johnson noise formula
once the high temperature limit of $S(\omega)$ is taken
and $R(\omega)$ is identified with the resistance of the sample.
In mesoscopics, the phase of the electron wavefunction
is changed by a time dependent potential
$V(x(t),t)$ by an amount given by the electrostatic
Aharonov-Bohm effect, $\delta\phi = (e/\hbar)\int Vdt$.
Following Stern {\it et al.} \cite{sai}, we define dephasing 
such that $\langle(\delta\phi)^2\rangle \sim 1$, which  
occurs over a timescale of $\tau_\phi$. The dephasing
time can be computed from

\begin{equation}
1\simeq \langle(\delta\phi)^2\rangle ={e^2 \over \hbar^2}\int_0^{\tau_\phi}
\int_0^{\tau_\phi}dt dt^\prime \langle V(t)V(t^\prime)\rangle.
\end{equation}
\noindent
Most of the contribution to $\langle(\delta\phi)^2\rangle$ will occur when
$\omega\tau_\phi < 1$ \cite{sai} and the environment can essentially
be considered to be stationary over the time $\tau_\phi$.
Eq. 9 can then be approximated by $\langle V^2 \rangle=\hbar^2/e^2\tau_\phi^2$ 
becoming the left hand side of Eq. 8. The next important step
is to remember that in mesoscopics, the conductance $G$ of a phase
coherent volume fluctuates as the interference of all paths
occurs at every point in the phase coherent volume
by an amount given by universal conductance fluctuation theory
\cite{stone}, $\Delta G= e^2/h = \Delta R_\phi/R_\phi^2$ on average.
Therefore we believe the parameter which controls the irreversibility
in Eq. 8 is not the sample resistance $R$ but rather the fluctuation
$\Delta R_\phi = e^2 R_\phi^2/h$. Since we are interested
in the intrinsic decoherence time, we take the low temperature
limit, $S(\omega) \rightarrow \hbar\omega/2, \tau_\phi \rightarrow
\tau_0$, and assume that $R_\phi(\omega)$ is frequency independent
over the same limits of integration used in Eq. 5. Just as in
the derivation of Eq. 6, we use $R/L = R_\phi/L_\phi$ and we
easily arrive at the same result as given in Eq. 6.

\par

In conclusion, we have shown that the phase coherence time in
mesoscopic systems does not go to infinity as predicted by most  
theories but will always be limited by the zero-point
fluctuations of the intrinsic electromagnetic environment.
We have shown that the ubiquitous saturation
of $\tau_\phi$ found in all low temperature experiments
on 1D mesoscopic systems can be quantitatively understood
by incorporating the zero-point fluctuations into    
previous electron-electron interaction theories. 
We have demonstrated how the complete temperature 
dependence of $\tau_\phi$ can be calculated using the
full spectral density function for three examples, long
metal wires, long semiconductor wires and completely phase 
coherent short wires. This functional form appears to be
consistent with data from most experiments. In addition,     
we predict the magnitude 
of the low temperature saturation value of this decoherence time  
from the knowledge of only $R/L$ and the classical diffusion
constant in agreement with most experiments.  

\par
We thank A. Raval and J. Schwarz for useful discussions.
This work is supported by the NSF under  
contract No. DMR9510416.

\narrowtext

\end{document}